# Isomorphism between the Peres and Penrose proofs of the BKS theorem in three dimensions

Elizabeth Gould and P.K.Aravind
Physics Department
Worcester Polytechnic Institute
Worcester, MA 01609
paravind@wpi.edu

ABSTRACT

It is shown that the 33 complex rays in three dimensions used by Penrose to prove the Bell-Kochen-Specker theorem have the same orthogonality relations as the 33 real rays of Peres, and therefore provide an isomorphic proof of the theorem. It is further shown that the Peres and Penrose rays are just two members of a continuous three-parameter family of unitarily inequivalent rays that prove the theorem.

Some time back Peres [1] gave a proof of the Bell-Kochen-Specker (BKS) theorem [2] using 33 real rays (or directions) in three dimensions. Penrose [3] later gave a different proof of the theorem using 33 complex rays in three dimensions. Penrose pointed out that his set of rays is essentially complex (i.e., there is no basis in which the components of all the rays can be made real) and that there is no Hilbert space rotation that will take his rays into those of Peres. It might therefore be thought that the proofs of the BKS theorem based on the two sets of rays are essentially different. However we show in this paper that the Kochen-Specker diagrams of the Peres and Penrose rays are identical, and that they therefore furnish isomorphic proofs of the BKS theorem. We exploit the common cubic symmetry of both sets of rays to give a unified proof of the BKS theorem for them that is shorter than the one given by Peres. Finally, we demonstrate that the Peres and Penrose rays are just two members of a continuous three-parameter family of unitarily inequivalent rays that prove the theorem.

The 33 rays of Peres can be visualized in terms of the geometry of a cube. Each ray goes out from the center of a cube to a point on its surface as follows: three rays (numbered 1-3) go to the midpoints of the faces, six (numbered 4-9) to the midpoints of the edges, twelve (numbered 10-21) to the midpoints of the edges of squares inscribed in the faces, and twelve more (numbered 22-33) to the vertices of the preceding squares. Choosing a Cartesian coordinate system with its origin at the center of the cube and its axes parallel to the cube edges, the components of the rays (up to an overall multiplicative factor) are

$$\begin{aligned}
&1 = 100 &&2 = 010 &&3 = 001 &&4 = 011 &&5 = 01\bar{1} &&6 = 101 \\
&7 = 10\bar{1} &&8 = 1\bar{1}0 &&9 = 110 &&10 = 2\bar{1}1 &&11 = 211 &&12 = 2\bar{1}\bar{1} \\
&13 = 21\bar{1} &&14 = \bar{1}21 &&15 = 121 &&16 = \bar{1}2\bar{1} &&17 = 12\bar{1} &&18 = 112 \\
&19 = \bar{1}12 &&20 = 1\bar{1}2 &&21 = \bar{1}\bar{1}2 &&22 = 102 &&23 = \bar{1}20 &&24 = 120 \\
&25 = \bar{1}02 &&26 = 012 &&27 = 2\bar{1}0 &&28 = 210 &&29 = 0\bar{1}2 &&30 = 021 \\
&31 = 201 &&32 = 20\bar{1} &&33 = 02\bar{1}
\end{aligned} \quad (1)$$

where we have omitted commas between components, put a bar over a number to indicate its negative and written $\sqrt{2}$ as 2 everywhere for typographic simplicity. The orthogonality table (or "Kochen-Specker" diagram) of the above rays is easily constructed and seen to consist of the 16 triads and 24 dyads of mutually orthogonal rays listed in Table 1.

| 1 2 3 | 1 4 5 | 1 26 33 | 1 29 30 | 2 6 7 | 2 22 32 | 2 25 31 | 3 8 9 |
|---|---|---|---|---|---|---|---|
| 3 23 28 | 3 24 27 | 4 10 13 | 5 11 12 | 6 14 17 | 7 15 16 | 8 18 21 | 9 19 20 |

| 10 24 | 10 25 | 11 23 | 11 25 | 12 22 | 12 24 | 13 22 | 13 23 |
|---|---|---|---|---|---|---|---|
| 14 28 | 14 29 | 15 27 | 15 29 | 16 26 | 16 28 | 17 26 | 17 27 |
| 18 32 | 18 33 | 19 31 | 19 33 | 20 30 | 20 32 | 21 30 | 21 31 |

TABLE 1. Common orthogonality table of the Peres and Penrose rays, consisting of 16 triads (at the top) and 24 dyads (at the bottom).

The Penrose rays are also related to the geometry of a cube. Each ray is most simply specified in terms of its two Majorana vectors (see [4]-[6] for a review of the Majorana description of spin systems). Consider the 18 vectors from the center of a cube to the midpoints of its faces and edges. The Majorana vectors (or M-vectors) of the Penrose rays can be obtained by pairing up these 18 vectors as follows:

$$\begin{aligned}
&1 = 100, \bar{1}00 &&2 = 010, 0\bar{1}0 &&3 = 001, 00\bar{1} &&4 = 011, 0\bar{1}\bar{1} &&5 = 01\bar{1}, 0\bar{1}1 \\
&6 = 101, \bar{1}0\bar{1} &&7 = 10\bar{1}, \bar{1}01 &&8 = 110, \bar{1}\bar{1}0 &&9 = 1\bar{1}0, \bar{1}10 &&10 = 011, 011 \\
&11 = 01\bar{1}, 01\bar{1} &&12 = 0\bar{1}1, 0\bar{1}1 &&13 = 0\bar{1}\bar{1}, 0\bar{1}\bar{1} &&14 = 101, 101 &&15 = 10\bar{1}, 10\bar{1} \\
&16 = \bar{1}01, \bar{1}01 &&17 = \bar{1}0\bar{1}, \bar{1}0\bar{1} &&18 = 110, 110 &&19 = 1\bar{1}0, 1\bar{1}0 &&20 = \bar{1}10, \bar{1}10 \\
&21 = \bar{1}\bar{1}0, \bar{1}\bar{1}0 &&22 = 011, 01\bar{1} &&23 = 011, 0\bar{1}1 &&24 = 0\bar{1}\bar{1}, 01\bar{1} &&25 = 0\bar{1}\bar{1}, 0\bar{1}1 \\
&26 = 101, 10\bar{1} &&27 = 101, \bar{1}01 &&28 = \bar{1}0\bar{1}, 10\bar{1} &&29 = \bar{1}0\bar{1}, \bar{1}01 &&30 = 110, 1\bar{1}0 \\
&31 = 110, \bar{1}10 &&32 = \bar{1}\bar{1}0, 1\bar{1}0 &&33 = \bar{1}\bar{1}0, \bar{1}10
\end{aligned} \quad (2)$$

As in (1), we have omitted commas between the components of the vectors above and used a bar over a number to indicate its negative. Also, we have not written the M-vectors as unit vectors (which they really are) but multiplied them by suitable factors to make their components simple integers. Like the Peres rays, the Penrose rays fall into four distinct classes of 3,6,12 and 12 members consisting of rays 1-3, 4-9, 10-21 and 22-33, respectively. The difference between the classes can again be understood with reference to the geometry of a cube: the M-vectors of the rays in the first class point from the center of the cube to the midpoints of opposite faces, those in the second class to the midpoints of opposite edges, those in the third class (which consists of pairs of identical vectors) to the midpoints of the edges, and those in the last class to the midpoints of edges that are opposite across a face. The common cubic symmetry of the Peres and Penrose rays, as well as their common class structure, hints at the existence of a deeper connection between them.

To uncover this connection, we need to work out the orthogonalities of the Penrose rays. One way of doing this is to use the criteria for the orthogonality of spin-1 states in terms of their M-vectors given by Penrose [3]. Another way is to note that the squared modulus of the inner product of the rays $|\vec{a}_1, \vec{a}_2\rangle$ and $|\vec{b}_1, \vec{b}_2\rangle$ (where both are labeled by their M-vectors) is

$$\left|\langle \vec{b}_1, \vec{b}_2 | \vec{a}_1, \vec{a}_2 \rangle\right|^2 = \frac{2\left[\left(1+\vec{a}_1\cdot\vec{b}_1\right)\left(1+\vec{a}_2\cdot\vec{b}_2\right)+\left(1+\vec{a}_1\cdot\vec{b}_2\right)\left(1+\vec{a}_2\cdot\vec{b}_1\right)\right]-\left(1-\vec{a}_1\cdot\vec{a}_2\right)\left(1-\vec{b}_1\cdot\vec{b}_2\right)}{\left(3+\vec{a}_1\cdot\vec{a}_2\right)\left(3+\vec{b}_1\cdot\vec{b}_2\right)} \quad . \quad (3)$$

Using either of these methods, one finds that the orthogonalities of the Penrose rays are identical to those of the Peres rays shown in Table 1. The numbering schemes of the two sets of rays were chosen expressly to ensure this convergence. We have thus established our first result, namely, the identity of the Kochen-Specker diagrams of the Peres and Penrose rays.

It is easily seen, as noted by Penrose, that the Peres and Penrose rays are not unitarily equivalent. The magnitude of the inner product of rays 9 and 14 of the Peres set is $(2-\sqrt{2})/4$, whereas it is $\sqrt{6}/4$ for the same rays of the Penrose set. This and other examples demonstrate that the isomorphism of Table 1 is confined to the orthogonalities alone and does not extend to the unitary equivalence of the two sets of rays.

We come next to our second point, which is to give a unified proof of the BKS theorem for both the Peres and Penrose rays. The cubic symmetry of the rays allows the proof to be given in the seven steps shown in Table 2, with the caption below the table conveying the essential idea of the proof and the following paragraph providing a fuller explanation. It should be stressed that the entire proof, including the symmetry operations in the first two steps, holds equally for the Penrose and Peres rays.

|       |       |       |       |       |       |                                       |
|-------|-------|-------|-------|-------|-------|---------------------------------------|
| **1** | 2     | 3     |       |       |       | (120° rotation about 111 direction)   |
| 4     | **10**| 13    | 5     | **11**| 12    | (90°,180° or 270° rotation about x-axis) |
| 2     | 25    | **31**|       |       |       |                                       |
| 3     | 24    | **27**|       |       |       |                                       |
| 3     | 23    | **28**|       |       |       |                                       |
| **6** | 14    | 17    |       |       |       |                                       |
| 7     | 15    | 16    |       |       |       |                                       |

TABLE 2. A common non-coloring proof of the BKS theorem for the Penrose and Peres rays. Each line shows a triad of mutually orthogonal rays, with green rays shown in boldface and red rays in ordinary type. Rays that have their colors assigned as a matter of choice are underlined, while those whose colors are forced are not. The choice of the green rays in the first two steps is arbitrary, but the indicated symmetries allow all the alternative choices to be mapped into the one picked, making the consideration of the alternatives unnecessary. The colors of all the rays listed after the second step are forced and lead to the totally red triad at the end, showing the impossibility of a viable coloring.

     A proof of the BKS theorem requires showing, in the language of Peres [1], that it is impossible to color each of the rays red or green in such a way that there is exactly one green ray (and two red rays) in every triad and at most one green ray in every dyad. We carry out the proof by contradiction, by showing that all possible attempts at constructing a satisfactory coloring end in failure. Consider the triad 1 2 3. If a viable coloring exists, one of the rays in it must be green. We can choose the green ray to be 1 without any loss of generality because either ray 2 or ray 3 can be made to pass into 1 by a rotation about the 111 direction that leaves the system of rays invariant as a whole. Coloring ray 1 green forces rays 2,3,4,5,26,33,29 and 30 (all of which are orthogonal to 1) to be red. Let us next consider the triads 4 10 13 and 5 11 12. Since 4 and 5 are already red, both members of one of the pairs (10,11), (10,12), (13,12) or (11,13) must be colored green. The pair (10,11) has been picked in Table 2, but this involves no loss of generality since a rotation by 90°,180° or 270° about the x-axis can be used to replace it by (12,10), (13,12) or (11,13), respectively, while keeping the green ray 1 fixed and permuting the eight red rays among themselves. With these choices made, the colors of the rest of the rays in the table are forced and lead to the completely red triad shown. This demonstrates that no viable coloring of the rays is possible and proves the BKS theorem. It is worth noting that this proof requires only 7 rays to be colored green, in contrast to the 10 required in Peres' proof [1].

Both the Penrose and Peres rays form a critical set, in that deleting even a single ray from either of them causes the BKS proof to fail. To demonstrate criticality, it suffices to show that deletion of a single ray from each of the four symmetry classes noted produces a colorable set. This is easily done. For example, deleting ray 1 from the first class and coloring rays 2,4,8,12,14, 16, 19,23 and 27 green and all the others red produces a satisfactory coloring. Similar demonstrations can be given for the other three classes.

Finally, we note that the Peres and Penrose rays are not the only ones obeying the orthogonality relations of Table 1. Table 3 shows the most general family of rays obeying these relations. These rays involve three complex parameters $a, b$ and $c$ whose amplitudes are fixed but whose phases are free to vary. The Peres and Penrose rays are special cases of this family for $a = 1, b = 1, c = \sqrt{2}$ and $a = -i, b = -1, c = -\sqrt{2}$, respectively. In the case of the Penrose rays one must apply the rotation

$$\begin{pmatrix} 1 & 1 & 0 \\ 0 & 0 & \sqrt{2} \\ -1 & 1 & 0 \end{pmatrix}$$

to the rays in Table 3 before working out their M-vectors to recover the forms listed in (2). It should be stressed that the different members of this family of rays are not unitarily equivalent to each other, although they share a common pattern of orthogonalities.

| | | | | |
|---|---|---|---|---|
| $1 = (1,0,0)$ | $2 = (0,1,0)$ | $3 = (0,0,1)$ | $4 = (0,1,a)$ | $5 = (0,a^*,-1)$ |
| $6 = (1,0,b)$ | $7 = (b^*,0,-1)$ | $8 = (1,k,0)$ | $9 = (k^*,-1,0)$ | $10 = (a^*c^*,-a^*,1)$ |
| $11 = (c^*,1,a)$ | $12 = (-c^*,1,a)$ | $13 = (a^*c^*,a^*,-1)$ | $14 = (-b^*,b^*c,1)$ | $15 = (1,c,b)$ |
| $16 = (1,-c,b)$ | $17 = (b^*,b^*c,-1)$ | $18 = (-k^*,1,bc^*)$ | $19 = (1,k,-ac)$ | $20 = (1,k,ac)$ |
| $21 = (k^*,-1,bc^*)$ | $22 = (1,0,ac)$ | $23 = (1,-c,0)$ | $24 = (1,c,0)$ | $25 = (1,0,-ac)$ |
| $26 = (0,1,bc^*)$ | $27 = (-c^*,1,0)$ | $28 = (c^*,1,0)$ | $29 = (0,1,-bc^*)$ | $30 = (0,b^*c,1)$ |
| $31 = (a^*c^*,0,1)$ | $32 = (-a^*c^*,0,1)$ | $33 = (0,-b^*c,1)$ | | |

TABLE 3. The most general set of rays obeying the orthogonality relations of Table 1. $a, b$ and $c$ are arbitrary complex numbers such that $|a|^2 = |b|^2 = 1$, $|c|^2 = 2$, and $k = -ab^*c/c^*$.

Finally we note the existence of another set of 33 real rays in three dimensions, different from that of Peres, that also furnishes a proof of the BKS theorem. This set was constructed by Bub [7] based on a tautology proposed by K.Schutte. However the Schutte-Bub set of rays possesses a different Kochen-Specker diagram from the Peres and Penrose sets, and so does not fall into the general class found here.

In closing, we mention why getting further insights into proofs of the BKS theorem is of value. For one, BKS proofs serve as the basis for quantum key distribution protocols certified by quantum value indefiniteness [8,9]. For another, BKS proofs can be converted into inequalities for non-contextual hidden variable models, which can be experimentally tested [10-14]. A third reason is that these proofs are linked to applications of "quantum contextuality" such as random number generation, parity-oblivious transfer and multiplexing tasks [15,16]. The connection between fundamental theory and applications is well illustrated by progress in this area of research.

**Acknowledgement.** We thank Mordecai Waegell for several useful discussions on the subject of this paper.